# Measurement Error Effects of Beam Parameters Determined by Beam Profiles

**Ji-Ho Jang, Hyo Jae Jang and Dong-O Jeon**

*RISP, Institute for Basic Science, Daejeon 305-811, Republic of Korea*

A conventional method to determine beam parameters is using the profile measurements and converting them into the values of twiss parameters and beam emittance at a specified position. The beam information can be used to improve transverse beam matching between two different beam lines or accelerating structures. This work is related with the measurement error effects of the beam parameters and the optimal number of profile monitors in a section between MEBT (medium energy beam transport) and QWR (quarter wave resonator) of RAON linear accelerator.

Email: jhjang@ibs.re.kr

Fax: +82-42-878-8799

# I. INTRODUCTION

The RAON diver linac of RISP (rare isotope science project) consists of an injector with ECR (electron cyclotron resonance) ion source and RFQ (radio frequency quadrupoles), and an SCL (superconducting linac) with QWR (quarter-wave resonator), HWR (half-wave resonator) and SSR (single spoke resonator). The layout of the linac is shown in Figure 1. The reference particle is the two charge state uranium beams, $^{238}U^{33+,34+}$ with the kinetic energy of 10 keV/u from the ion source. The RFQ can accelerate heavy ion beams up to 500 keV/u and the beam power reaches 400 kW after SCL with the kinetic energy of 200 MeV/u [1, 2].

The transverse matching between different accelerating structures or beam lines is important issue of beam optics in order to minimize the emittance growth and reduce beam loss along the accelerator [3]. The beam parameters should be determined at the entrance of matching qaudrupoles and set the field gradient values of the quadrupole magnets to achieve the beam matching into the next lattice of the linac. A method is using field profile measurement to obtain the beam parameters at the specified position [4]. The rms (root mean square) beam sizes obtained by the profile monitors at several positions are converted into the information of the beam parameters, twiss parameters and beam emittance at the position.

In RAON linac, beam profile monitors will be installed in the initial region of each different accelerating structure. They can be used to measure beam profiles and match the ion beams into the structure in the transverse directions. This work is related with how the errors in the profile measurement affect the determined beam parameters. We focused on the region of the MEBT (medium energy beam transport) of the injector and the QWR section of SCL in this paper. From this study, we can obtain the relation between errors of the measurement and the parameters, and determine the optimized number of profile measurement systems.

## II. THEORY

The beta function after an interesting region of accelerator or beam lines can be obtained by using the transfer matrix elements in the region as follows [3],

$$\beta_f = \begin{pmatrix} R_{11}^2 & -2R_{11}R_{12} & R_{12}^2 \end{pmatrix} \begin{pmatrix} \beta_i \\ \alpha_i \\ \gamma_i \end{pmatrix}, \quad (1)$$

where $\alpha$, $\beta$, $\gamma$ represent twiss parameters with the subscripts, $i$ for the positon where beam parameters are determined, and $f$ for the position where the beam profiles are measured. The parameters $R_{ij}$ are the components of 2×2 transfer matrix for the region. By multiplying beam emittance and considering $n$ different positions of profile measurement, we can obtain

$$\begin{pmatrix} \left(\sigma^{(1)}\right)^2 \\ \left(\sigma^{(2)}\right)^2 \\ \left(\sigma^{(3)}\right)^2 \\ \ldots \\ \left(\sigma^{(n)}\right)^2 \end{pmatrix} = \begin{pmatrix} \left(R_{11}^{(1)}\right)^2 & -2R_{11}^{(1)}R_{12}^{(1)} & \left(R_{12}^{(1)}\right)^2 \\ \left(R_{11}^{(2)}\right)^2 & -2R_{11}^{(2)}R_{12}^{(2)} & \left(R_{12}^{(2)}\right)^2 \\ \left(R_{11}^{(3)}\right)^2 & -2R_{11}^{(3)}R_{12}^{(3)} & \left(R_{12}^{(3)}\right)^2 \\ & \ldots & \\ \left(R_{11}^{(n)}\right)^2 & -2R_{11}^{(n)}R_{12}^{(n)} & \left(R_{12}^{(n)}\right)^2 \end{pmatrix} \begin{pmatrix} \varepsilon\beta_i \\ \varepsilon\alpha_i \\ \varepsilon\gamma_i \end{pmatrix}, \quad (2)$$

where $\sigma^{(i)} = \sqrt{\varepsilon_f^{(i)}\beta_f^{(i)}}$ is the rms beam size when $\varepsilon$ is a rms emittance and it is measured by $i$-th beam profile monitor. This equation can be symbolically expressed by using the corresponding vectors, $\Sigma$ and $b$, and the matrix, $M$ as follows,

$$\Sigma = M \cdot b, \quad (3)$$

where $M$ is $n \times 3$ matrix, $\Sigma$ is the vector of rms beam sizes with $n$ components, and $b$ represents the vector of twiss parameter times beam emittance with 3 components. The beam vector, $b$ can be determined by minimizing $\chi^2$ which is defined as follows,

$$\chi^2 = \sum_{k=1}^{n} \left( \Sigma_k - \sum_{i=1}^{3} M_{ki} b_i \right). \quad (4)$$

Then the twiss parameter, $\alpha_i$ and $\beta_i$, and beam emittance, $\varepsilon_i$, can be determined by the equation,

$$\begin{aligned} \varepsilon_i &= \sqrt{b_1 b_3 - b_2^2}, \\ \beta_i &= b_1/\varepsilon_i, \\ \alpha_i &= b_2/\varepsilon_i. \end{aligned} \quad (5)$$

where we used $\beta_i \gamma_i - \alpha_i^2 = 1$.

### III. RESULTS AND DISCUSSION

In this work, we studied beam parameter in the region with last four quadrupole magnets in MEBT for beam matching, and the initial 7 periods of double lattice of QWR. In order to determine the beam parameters at the entrance of the matching quadrupole set, we considered that beam profile monitors are located in the third to seventh period of QWR lattice.

The beam profiles in transverse directions are given in Figure 2 where the blue and red lines are the profiles in horizontal and vertical directions, respectively. It also shows the positions of the beam profile monitors in the warm sections between QWR cryomodules. We turned RF off in QWR cavities in this calculation. For the beam dynamics calculation, we used TRACK code with 500,000 macro particles in cases of the single charge states, 33 or 34. [5].

In order to determine the optimal number of profile monitors, we compared the rms beam size between the input values of the TRACK simulation and the determined values, $\sigma_i = \sqrt{\varepsilon_i \beta_i}$ by the method in section II. The measurement errors with $\sigma = 200$ μm in the measured rms beam size have the Gaussian distribution. We considered 3 different cases with 3, 4 and 5 beam profile monitors in order to study error sensitivity when the beam parameters in Eq. 5 is determined from the measurement data.

First of all, we tested how the input beam parameters, twiss parameter and beam emittance, can be accurately determined at the input point of Figure 2. Table 1 summarized the results of central values of the distribution of the rms beam sizes. We found that beam parameters obtained by the beam profiles become very similar to the simulation input values except a twiss parameter α in the case of charge state of 34. It gives about 35% deviation from the simulation result. In the following analysis we will use rms beam size to obtain the optimal number of profile monitors.

Figure 3 and Figure 4 show the distribution of the rms beam sizes for charge state of 33 and 34, respectively. In each figure, there are 3 cases with different number of profile monitors. The figures also include the Gaussian fitting results for the distributions. Table 2 includes the standard deviations of the Gaussian fitting results. It indicates the relative accuracy of determined beam size between the cases with different number of beam profile monitors. In the case of 3 profile monitors, we found that the determined values of beam profile include relatively large errors in every case of different charge states. The errors should depend on the measurement errors. When we use 4 profile monitors, errors in the determined rms beam sizes are reduced about 40% from the case of 3 monitors. We also found that the difference between 4 and 5 profile monitors becomes relatively small. Then we can conclude that the optimal number of beam profile monitors is 4 for practical purposes.

Finally, we calculated the standard deviation of the distribution of rms beam sizes depending on the standard deviation of the error distribution in beam profile measurement with 4 beam profile monitors. Figure 5 shows the result which includes the linear fit of the data. This shows that the beam size error is linearly proportional to the measurement error of the beam profile. For example the allowed profile measurement error should be $\sigma < 110$ μm in the profile measurement in order to achieve the beam size error less than 100 μm.

### III. CONCLUSION

We studied the beam profile measurement error effects on the beam parameters at the entrance part of the matching quadruples in RAON MEBT. From this study we found that the optimal number of beam profile monitors is 4 for practical purposes to determine beam parameters. We also studied the relation between the errors of the profile measurement and the errors of beam parameters determined by profile measurement method.

### ACKNOWLEDGEMENT

This work was supported by the Rare Isotope Science Project of Institute for Basic Science funded by Ministry of Science, ICT and Future Planning.

Table 1. The beam parameters determined by using 3, 4 and 5 beam profile monitors (BPM) for single charge states of 33 or 34.

| Parameters | BPM Numbers | 33 | 34 |
|---|---|---|---|
| Unnormalized rms Emittance [π mm-mrad] | Simulation | 3.580 | 3.579 |
| | 3 | 3.580 | 3.577 |
| | 4 | 3.580 | 3.577 |
| | 5 | 3.580 | 3.579 |
| α | Simulation | -0.228 | -0.0227 |
| | 3 | -0.221 | -0.0152 |
| | 4 | -0.221 | -0.0151 |
| | 5 | -0.221 | -0.0147 |
| β [mm/mrad] | Simulation | 1.776 | 1.586 |
| | 3 | 1.770 | 1.585 |
| | 4 | 1.770 | 1.585 |
| | 5 | 1.770 | 1.584 |
| rms beam size [mm] | Simulation | 2.521 | 2.382 |
| | 3 | 2.517 | 2.381 |
| | 4 | 2.517 | 2.381 |
| | 5 | 2.517 | 2.381 |

Table 2. The standard deviation values of the distribution of the rms beam sizes with 3, 4 and 5 beam profile monitors (BPM) for single charge state of 33 or 34.

| BPM Number \ Charge states | 33 | 34 |
|---|---|---|
| 3 | 0.37 mm | 0.26 mm |
| 4 | 0.20 mm | 0.16 mm |
| 5 | 0.17 mm | 0.16 mm |

Fig. 1. Layout the RAON linear accelerator.

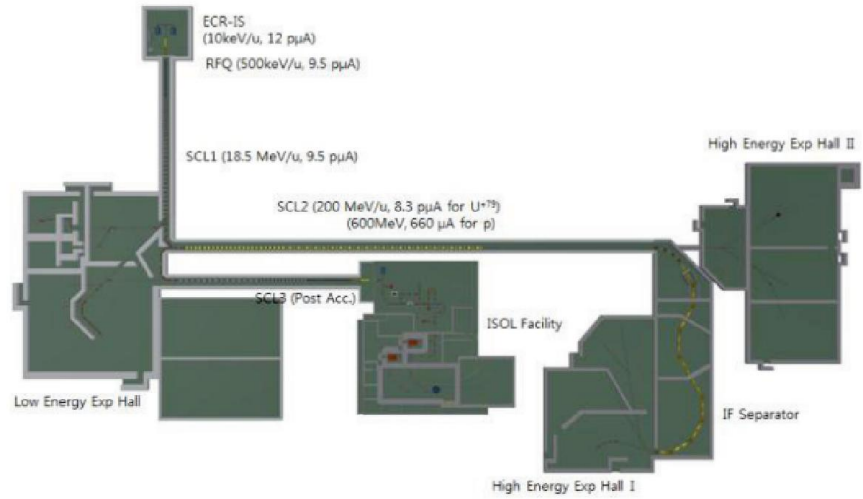

Fig. 2. TRACK simulation in MEBT and initial 7 periods of QWR lattice.

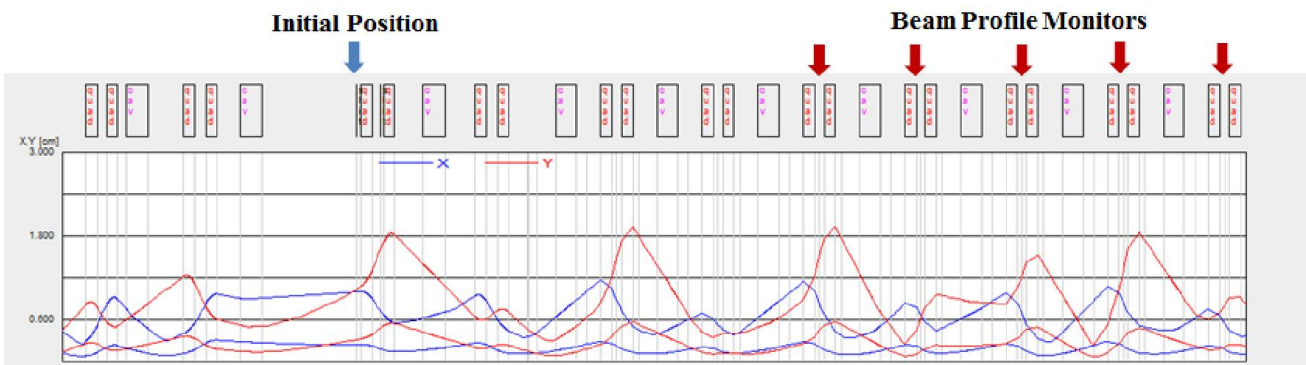

Fig. 3. Distribution of rms beam sizes and Gaussian fitting results for single charge state of 33 with (a) 3 beam profile monitors, (b) 4 beam profile monitors, and (c) 5 beam profile monitors.

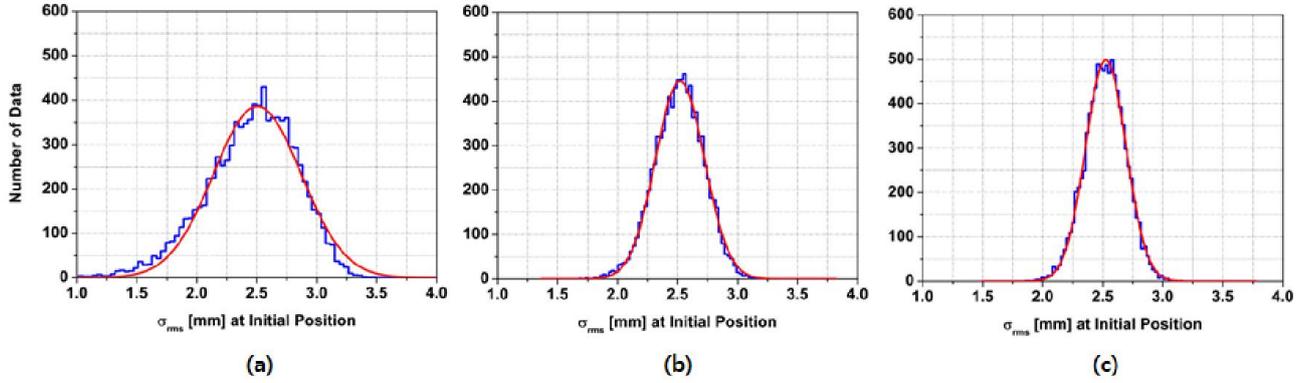

Fig. 4. Distribution of rms beam sizes and Gaussian fitting results for single charge state of 34 with (a) 3 beam profile monitors, (b) 4 beam profile monitors, and (c) 5 beam profile monitors.

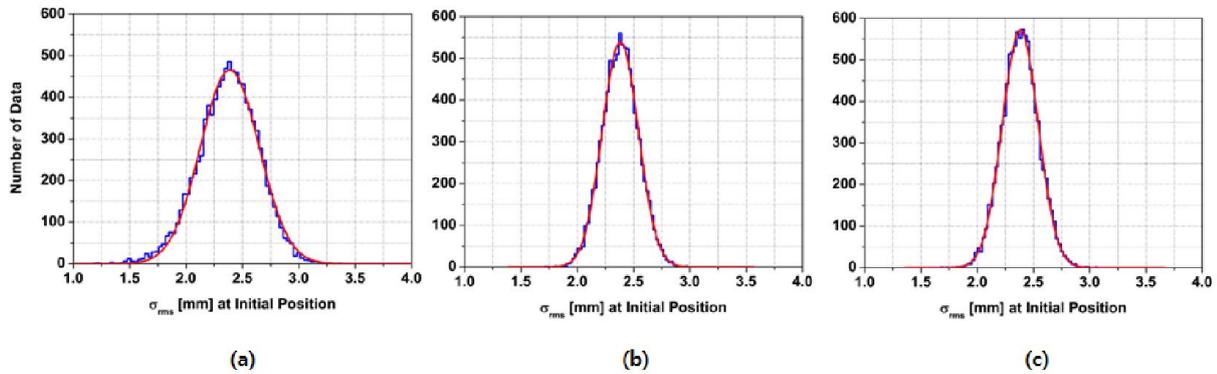

Fig. 5. The standard deviation of the rms beam size depending on the standard deviation of measurement errors in 4 beam profile monitor case.

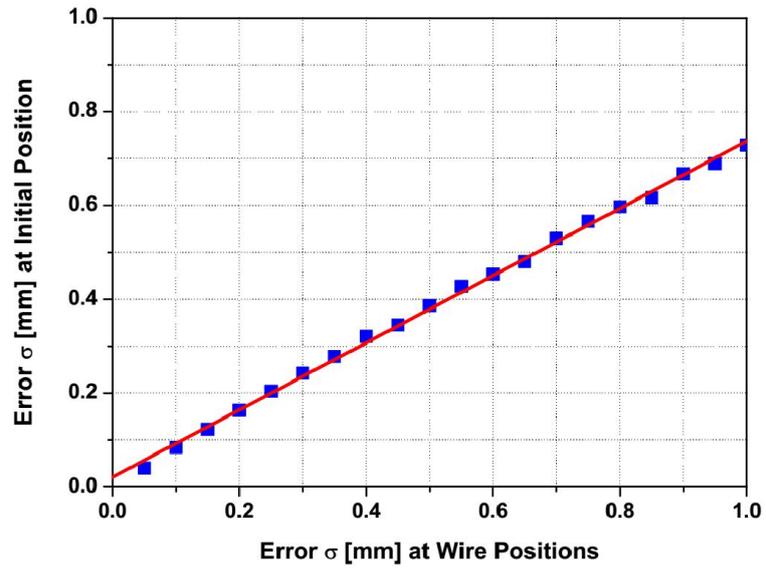